\documentclass[preprint,aps]{revtex4-1}

\usepackage{graphicx}
\usepackage{dcolumn}
\usepackage{bm}

\newcommand{\be}{\begin{equation}}
\newcommand{\ee}{\end{equation}}
\newcommand{\bdm}{\begin{displaymath}}
\newcommand{\edm}{\end{displaymath}}
\newcommand{\bea}{\begin{eqnarray*}}
\newcommand{\eea}{\end{eqnarray*}}

\begin{document}
\draft
\title{Suppression of the Melting Line in a Weakly
Disordered Flux-line System
}

\author{Gautam I. Menon}
\email{Email: menon@imsc.res.in}
\affiliation{The Institute of Mathematical Sciences, C.I.T. Campus, 
Taramani, Chennai 600 113, India}
\altaffiliation{Present address: Mechanobiology Institute and Department of Biological Sciences, National University of Singapore, 21 Lower Kent Ridge Road, Singapore 119077}
\author{G. Ravikumar}
\email{Email: gurazada@apsara.barc.ernet.in}
\affiliation{Technical Physics  Division,
Bhabha Atomic Research Centre, Mumbai 400 085, India}
\author{M.J. Higgins}
\email{Email:markhiggins58@gmail.com}
\affiliation{Princeton High School, Princeton NJ 08540, USA}
\author{S. Bhattacharya}
\email{Email:shobo@tifr.res.in}
\affiliation{Tata Institute of Fundamental Research, Homi Bhabha Road,
Colaba,Mumbai 400 005, India}
\date{\today}

\begin{abstract}

An analytic formula describing the suppression of
the equilibrium melting line by quenched point pinning
disorder is compared to data from ac susceptibility
and magnetization measurements in the mixed phase of
the layered dichalcogenide low-T$_c$ superconductor
2H-NbSe$_2$. This material exhibits a sharp peak
effect in the critical current j$_c$ close to the upper critical
field H$_{c2}(T)$.  Arguing that the disorder-suppressed 
melting line in this
system is to be identified with the locus of peak
positions of the critical current as magnetic
field and temperature are varied, we demonstrate that this formula
provides a  remarkably accurate fit to the experimental data over
three orders of magnitude in magnetic field.

\end{abstract}
\maketitle

The translational order of the Abrikosov flux line
lattice is disrupted both by thermal fluctuations and
by quenched disorder\cite{review1,natterman,shoboreview,baruch}.
Thermal fluctuations drive a first-order flux-lattice
melting transition in pure systems while quenched disorder
destabilizes translational long-range order even at zero
temperature\cite{nelson,larkin}.  A delicate balance between elastic
restoring forces  and the twin disordering effects of
thermal fluctuations and random pinning thus determines
flux line structure and phase behavior in the
mixed phase of type-II superconductors with quenched 
disorder\cite{shoboreview}.

Translational correlations cannot be truly long-ranged
in a crystal with quenched
disorder\cite{larkin}. Qualitatively different states may, however,
be identified through the nature of the 
decay of such correlations\cite{review1,natterman,shoboreview}. 
In the Bragg glass (BrG), a quasi-lattice 
phase believed to be stable at weak disorder and low temperatures, such
correlations  decay asymptotically
as power laws\cite{natterman1,giam}.  Upon increasing
either the effective disorder strength or the amplitude of
thermal fluctuations, the Bragg glass transforms into a phase in which such
correlations decay exponentially.  The disordered
liquid (DL) is one such stable state obtained at
large temperatures ($T$).  Another disordered state
is obtained in field ($H$) scans out of the Bragg
glass at low $T$\cite{shoboreview}. Such a disordered state has 
glassy attributes, including divergent time-scales for
structural relaxation and relatively short-ranged correlations. It may
represent a new thermodynamic phase, the vortex
glass phase, distinct from the equilibrium disordered 
liquid\cite{ffh, natterman,glass}.

\begin{figure}
\includegraphics[width=0.70\textwidth]{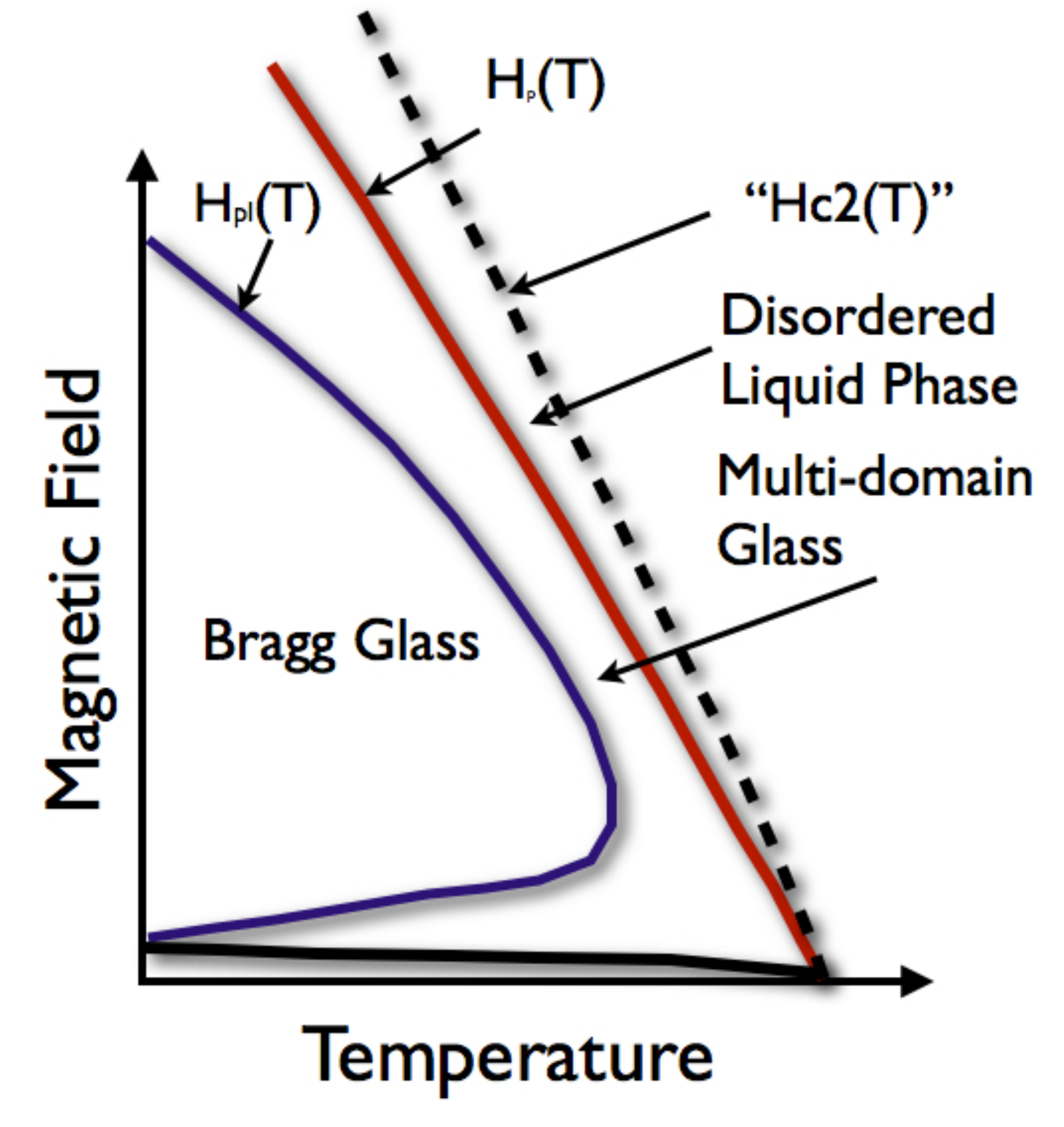}
\caption{\label{phasedia} [Color Online]  Generic phase diagram for type-II superconductors with point pinning disorder. The Bragg glass is
depicted as  melting directly into a ``multi-domain'' glass phase and only then into the disordered liquid upon increasing the temperature. The H$_{pl}$(T) (where $pl$ stands for ``plastic flow'' ) and H$_{p}$ lines, associated with the onset and the maximum of the peak effect anomaly in the ac susceptibility or critical current, are identified with the transitions between the Bragg glass and the multi-domain glass and the multi-domain glass and the disordered liquid respectively.  The boundary between Bragg Glass and multi-domain glass phase is reentrant\cite{nelson,ghosh}, although the regime of reentrance appears to be very small in experiments (it is exaggerated for clarity). The regime of intermediate vortex glass phase is also exaggerated in the figure vis a vis the experiments, where it appears as a very narrow sliver in the phase diagram, concomitant with other anomalies signaling a sequence of equilibrium phase transitions\cite{mypaper1}.}
\end{figure}

At large $H$, the DL phase appears to transform
continuously into the glass when $T$ is
reduced.  At low and intermediate values of
$H$, it appears to freeze discontinuously.  This freezing in
the intermediate interaction-dominated regime has
traditionally been understood as occurring directly
into the Bragg glass\cite{popular,kiervin}. 
Such a view is, however, at odds with a large body of
data in which anomalies associated with the
freezing transition, such as a discontinuity in
the magnetization, often appear well separated
from another structural transition which occurs
at lower temperatures, particularly for more
disordered samples\cite{mypaper1,mypaper2, mypaper3}. An
attractive way to reconcile a large number of experimental
observations on both high-T$_c$and low T$_c$ materials
is {\em via} the proposal that two transition
lines {\em always} separate the BrG phase from
the DL phase, with a sliver of disordered glassy phase
intervening. This possibility is illustrated in Fig.~\ref{phasedia}\cite{mypaper1,mypaper2,mypaper3},
with the intermediate glassy phase termed as a ``multi-domain glass''.
The phase diagram of Fig.~\ref{phasedia} has been  argued 
elsewhere to be a generic phase diagram for the mixed phase
with quenched point pinning\cite{mypaper3}. Among its distinctive
features is the identification of a smooth connection between high-field
and low-field glassy phases, a proposed equivalence of the peak effect as seen
in temperature scans as $T \rightarrow T_c$ and the ``fish-tail'' effect
seen in field scans at low temperatures, as well as a specific prediction for
vortex-line structure in the intermediate glassy phase\cite{mypaper1,mypaper2,mypaper3}.

In Fig.~\ref{phasedia}, the Bragg glass melts into the disordered
liquid phase on increasing $T$ via an intermediate
glassy phase\cite{mypaper1,mypaper2,mypaper3}. This
glassy regime broadens both at
large $H$ as well as very small $H$, reflecting
the increased importance of disorder both at high
fields, where  the multi-domain glass phase is encountered,
and at low fields, where a reentrant disordered
state has been predicted and indeed seen\cite{ghosh}.  The intervening sliver of
disordered phase has been identified with the peak
effect regime, the narrow region close to H$_{c2}$
in ($H,T$) space in which the critical current
$j_c$  {\em increases} anomalously in
a variety of low and high-T$_c$ superconductors\cite{shobo, peak1,peak2,peak3,peak4}.
Susceptibility measurements on a variety of low-T$_c$
materials indicate that the phenomenon of the peak
effect itself exhibits a generic two-step character,
in agreement with the proposed phase diagram.
In this picture, the first phase boundary encountered
when cooling from the disordered liquid state is
the remnant of the thermal melting line in the pure
system, renormalized suitably by disorder. This phase
boundary is then identified with the loci of 
peaks in $j_c$ in $(H,T)$ space. 

Significantly, this proposal  also motivates a new interpretation of
the classic problem of the peak effect: the peak effect
is simply a result of the disordered, high $j_c$ vortex
glass phase intervening between two relatively low $j_c$ phases,
the Bragg glass phase and the equilibrium disordered liquid.
 In this picture, the ``hardening''
of the vortex system as a consequence of the
transition into the multi-domain glass phase (rather
than the smooth ``softening'' envisaged in Pippard's
original scenario\cite{pippard} or a possibly sharp crossover from
collective to individual pinning\cite{mikitik,koopmann}), is the underlying cause of
the peak effect\cite{mypaper2}.  Such an increase in
critical currents in poly-domain structures has indeed been
seen in recent simulations\cite{moretti}. A recent study of this 
intermediate phase is reported in Ref.~\cite{pasquini}, while earlier work examined
this regime  using scanning ac Hall microscopy, proposing
the possibility of ``coexistence'' within the peak effect regime as an
explanation for the anomalies seen within this regime\cite{marchevsky}.

\begin{figure}
\includegraphics[width=0.65\textwidth]{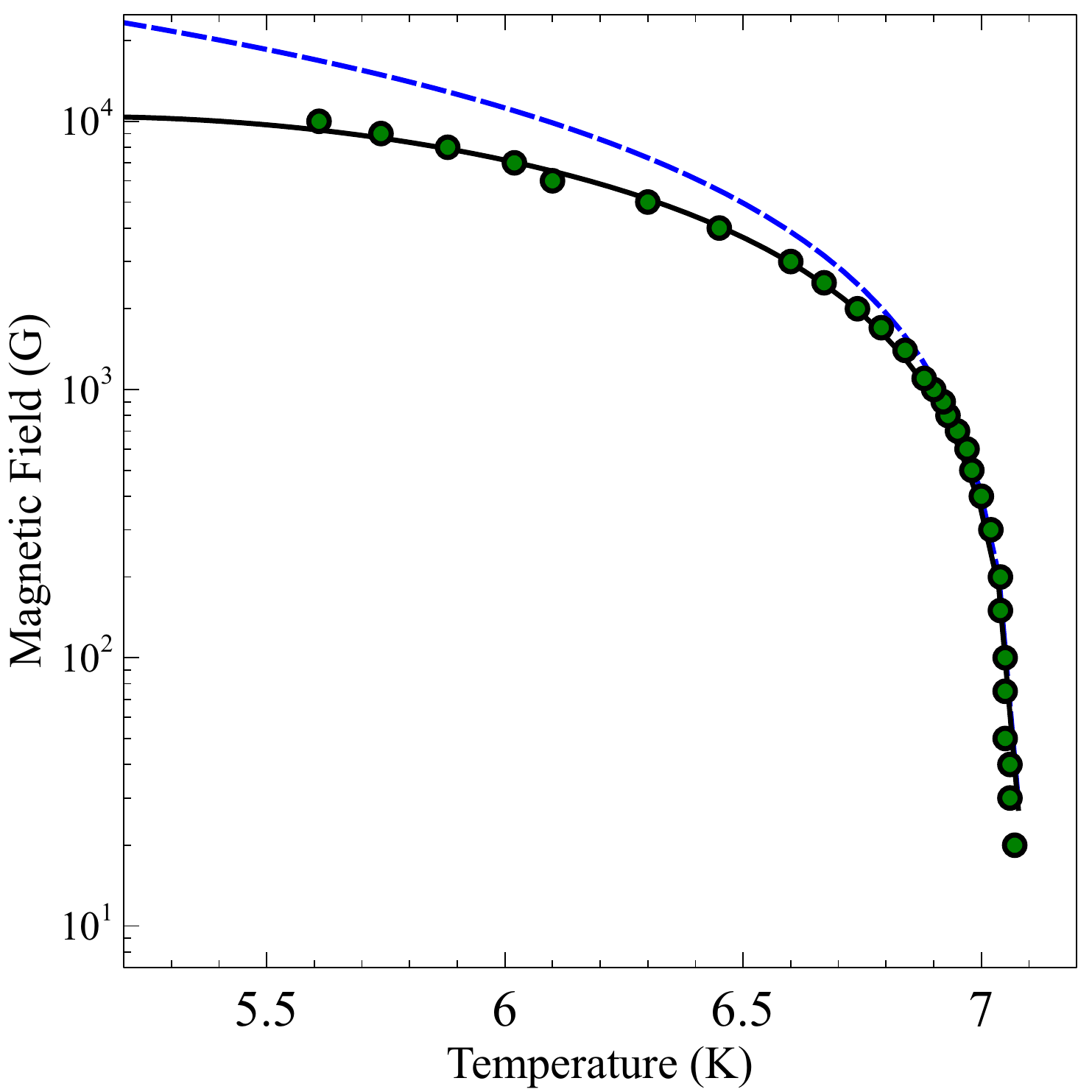}
\caption{\label{newdata} [Color Online] Experimental data (filled circles) showing the locus of points associated with the peak of the peak effect anomaly, as obtained from ac susceptibility measurements. The solid line passing through these point corresponds to the theoretical prediction in Eq.~\ref{eqvgdl} for the disorder-induced suppression of the phase boundary between the  multi-domain glass and the disordered liquid. The parameter values used in the fit are $C = 10^4$, $m = 10.0$, $T_c = 7.09$ and $\alpha =1.33$, assuming XY-type critical behaviour.
The dashed line represents a theoretical fit which corresponds to the pure system behavior, in which we set $m=0$.}
\end{figure}

While translational correlations in the BrG phase
are of power-law character, those in the DL phase
are short-ranged, with typical scales of order a few
interline spacings. We have  argued elsewhere
that the most appropriate description of structure
in the intermediate disordered phase is in terms of
a disordered arrangement of ordered domains, as in
a ``multi-domain glass'' (MG) phase\cite{mypaper1,mypaper2,mypaper3}.
Using formulae appropriate to the collective pinning
regime, a conjecture for typical domain sizes,
and experimentally obtained values of $j_c$ in very
weakly disordered samples of 2H-NbSe$_2$ indicate
$R_d/a \sim 10^6$, with $a$ the mean-intervortex
spacing, suggesting that the domain sizes in the MG
phase of systems with low levels of pinning can be
far larger than the correlation length at freezing
in the pure system\cite{mypaper1,mypaper2}.  Such a picture rationalizes the
association of magnetization discontinuities with the
transition out of the DL phase. Further, it suggests 
that examining the instability of the DL phase to
a phase with solid-like structural correlations may
be a good starting point for the calculation of the
DL-MG phase boundary.

\begin{figure}
\includegraphics[width=0.7\textwidth]{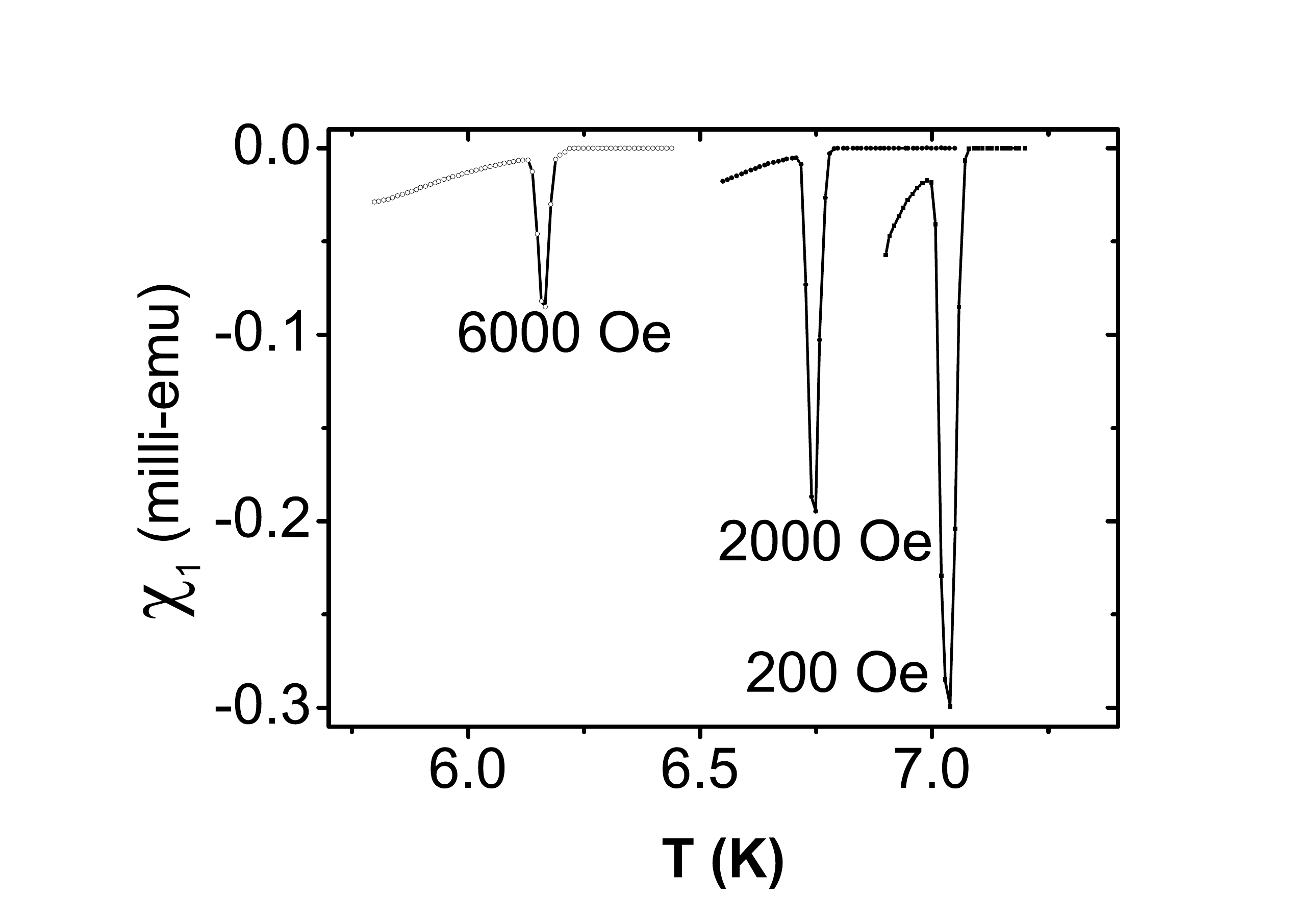}
\caption{\label{raviacchi} 
Real part of the ac susceptibility of an NbSe$_2$ crystal measured at three representative fields applied parallel to the 
c-axis. The data is measured using an ac field amplitude of 1 Oe and 10 Hz frequency.}
\end{figure}

This paper compares a generalized version
of the predictions of a simple semi-analytic
theory of the DL-MG phase transition, proposed 
in Ref.~\cite{mypaper2}, with results from
ac susceptibility and magnetization measurements on the low-T$_c$
dichalcogenide superconductor 2H-NbSe$_2$. This
material, with T$_c$ $\simeq$ 7.1K, shows a remarkably
sharp peak effect signal in an interval within about 10\%
of H$_{c2}(T)$. Structure in the peak effect regime in 2H-NbSe$_2$ has been
conclusively demonstrated to be domain-like,
validating the approach here\cite{fasano}; for related simulations see Ref.~\cite{moretti}. 

\begin{figure}
\includegraphics[width=0.7\textwidth]{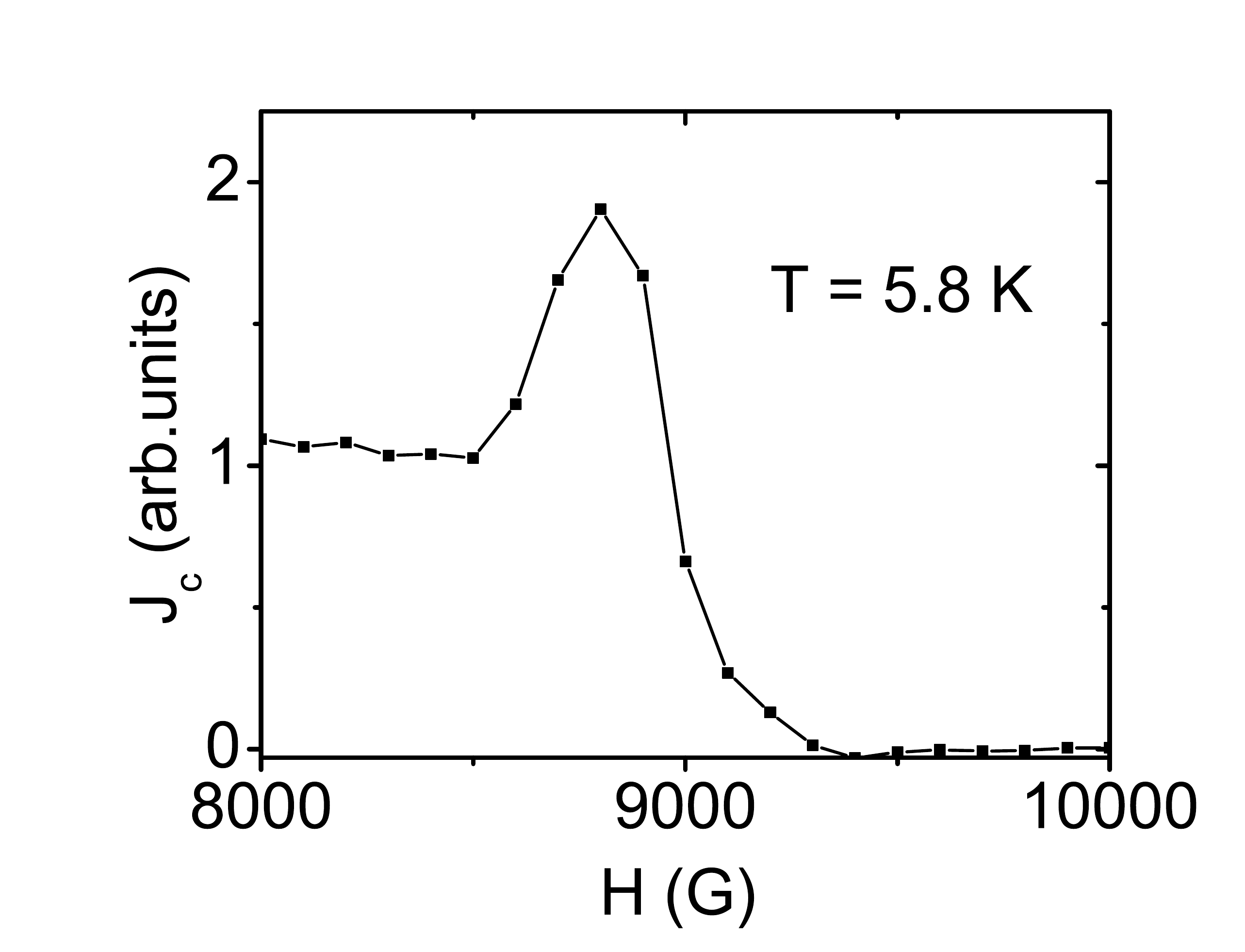}
\caption{\label{criticalcurrent} Critical current of the 2H-NbSe$_2$ sample as a function of applied field as obtained from dc magnetization measurements, illustrating the peak effect in this material, at a temperature of $T= 5.8 K$.}
\end{figure}

We find that a relatively simple analytic formula
provides a remarkably accurate parametrization of
the MG-DL transition line, the $H_p$ line in the phase diagram of Fig.~\ref{phasedia}, provided,
as argued extensively elsewhere, the loci of the peak
in $j_c$  is identified with this transition. This fit is shown in
Fig.~\ref{newdata}, using data from ac susceptibility measurements.  
 Fig.~\ref{raviacchi} exhibits 
representative data (see below) for the
ac susceptibility, while magnetization-based measurements of the critical
current are shown in Fig.~\ref{criticalcurrent} at one value of the temperature.
As can be 
seen from Fig.~\ref{newdata}, the fit to the data is of exceptional quality.

The approach of Ref.\cite{mypaper2}, described briefly below for completeness, draws from a
early replica theory of correlations in disordered
fluids and a replica generalization of the density
functional theory of freezing as applied originally to the
pancake vortex system in BSCCO\cite{gimcdg,dft}.
Such density functional  theories have been shown to provide a quantitatively
accurate picture of the freezing transition in the pure system,
including such details as the physics of the anomalous slope of the
melting line in these systems and a rationalization  of asymmetric hysteresis across the 
flux-lattice melting transition as induced
by the presence of free surfaces\cite{dft,gianni1,gianni2}.

Replica methods  are generalized to disordered fluids  through the model problem of
a  system of point particles interacting via a two-body
interaction and an explicit one-body disorder term
$V_d({\bf r})$\cite{gimcdg}. Here $V_d({\bf r})$ represents a
quenched, random, one-body potential, drawn from a
Gaussian distribution of zero mean and short ranged
correlations: $[V_d({\bf r})V_d({\bf r^\prime})] =
K(\mid{\bf r} - {\bf r}^\prime\mid)$, with $[\cdots]$
denoting an average over the disorder.

Two point correlation functions, within replica
theory, carry two replica indices, denoted by the
Greek letters $\alpha$ and $\beta$.  Thus, the
pair correlation functions $h^{\alpha\beta}(r)$
and the related direct correlation functions
$C^{\alpha\beta}(r)$ characterize the replicated
system at the level of two-particle correlations.
Assuming replica symmetry, as  appropriate to
the equilibrated disordered fluid, we define
the replica off-diagonal ($\alpha \neq
\beta$) and replica diagonal ($\alpha = \beta$)
correlation functions, writing $C^{\alpha\beta}
= C^{(1)}\delta_{\alpha\beta} + C^{(2)}(1 -
\delta_{\alpha\beta})$ and $h^{\alpha\beta}
= h^{(1)}\delta_{\alpha\beta} + h^{(2)}(1 -
\delta_{\alpha\beta})$.  Physically, the function
$h^{(1)}$ describes the disorder-averaged equal-time
(equilibrium) correlation of fluctuations of the
local density, whereas $h^{(2)}$ represents the
disorder-averaged correlation of disorder-induced
deviations of the time-averaged local density from
its average value $\rho_\ell$. Direct simulation tests of the predictions 
of the theoretical framework  for these correlation functions are reported 
in Ref.~\cite{ankush}.

The density functional theory of freezing takes such
correlations as input to a mean-field determination of
when the fluid, with density $\rho_\ell$  becomes first unstable to the formation
of a static density inhomogeneity. Applying replica
analysis  leads to a  density functional which qualitatively resembles the
density functional of a {\em pure} system, but with
renormalized correlations\cite{gimcdg}:
\begin{widetext}
\begin{eqnarray}
\frac{\Delta \Omega}{k_B T}&=&\int d{\bf r}
\left[\rho({\bf r})\ln \frac{\rho({\bf
r})}{\rho_\ell} - \delta \rho({\bf
r})\right] \nonumber \\
&&-\frac{1}{2}\int d{\bf r} \int d{\bf r}^\prime [C^{(1)}(\mid
{\bf r}-{\bf r}^\prime \mid) - C^{(2)}(\mid {\bf r}-{\bf r}^\prime \mid)]
[\rho({\bf r})-\rho_\ell][\rho({\bf
r^\prime})-\rho_\ell] + \ldots.
\label{eq312}
\end{eqnarray}
\end{widetext}
Assuming $\rho^\alpha({\bf r}) =
\rho({\bf r})$ for all $\alpha$, the
properties of the density functional are governed
by an {\em effective} direct correlation function
given by $C^{eff}(\mid {\bf r} - {\bf r}^\prime
\mid) = C^{(1)}(\mid {\bf r}-{\bf r}^\prime \mid)
- C^{(2)} (\mid {\bf r}-{\bf r}^\prime \mid)$. In
this description, disorder enters through (i) the
suppression of $C^{(1)}(\mid {\bf r}-{\bf r}^\prime
\mid)$ and (ii) the non-trivial character
of $C^{(2)} (\mid {\bf r}-{\bf r}^\prime \mid)$.

The inter-replica interaction $\beta
K(\rho,nd)$ appropriate to vortex lines and
pancake vortices is obtained assuming the principal
source of disorder to be atomic scale pinning
centers~\cite{chudnovsky}. A model calculation
yields $\beta V^{(2)}(\rho,nd) = -\beta K(\rho,nd)
\simeq -\Gamma^\prime\exp(-\rho^2/\xi^2)\delta_{n,0}$,
where $\beta V^{(2)}(\rho,nd) = \beta
V^{\alpha\beta}(\rho,nd)$ with $\alpha \neq \beta$.
$\xi \simeq 15 \AA$ is the coherence length
in the $ab$ plane, and $\Gamma^\prime \approx
10^{-5}\Gamma^2$ for point pinning of strength
$dr_0^2H_c^2/8\pi$, with $d$ the interlayer spacing
($\sim 15\AA$) for a layered superconductor (we
set $d = \xi$ for an isotropic superconductor),
$\Gamma = \beta d \Phi^2_0/4 \pi \lambda^2\/$ and
$\beta=1/k_B T\/$.  Defect densities of the order
of $10^{20}$/cm$^3$ are assumed\cite{gimcdg}. We will use only
the dependence of $\Gamma^\prime$ on the field and
temperature in our discussion i.e. $\Gamma^\prime \sim \Gamma^2 \sim 1/T^2$.

Further progress requires a  calculation of
the correlation functions $C^{(1)}(r)$ and $C^{(2)}(r)$.
A replica generalization of the Ornstein-Zernike
equations coupling $C^{(1)}(r)$ and $C^{(2)}(r)$ to
$h^{(1)})(r)$ and $h^{(2)}(r)$ can be derived: these
equations must be supplemented, as usual, with
appropriate closure schemes, such as the hypernetted
chain (HNC) or Percus Yevick (PY)\cite{hanmac,gimcdg}. 

In mean-field theory, the freezing transition of
the pure system occurs when the density functional
supports periodic solutions with a free energy lower
than that of the uniform fluid\cite{ry}.  We
now specialize to the flux-line lattice, in which case
properties of the freezing transition are controlled
by a two-dimensional $C^{eff}(q_\perp, q_z=0)$, the
Fourier transform of $C^{eff}(r_\perp,z)$. Setting
$q_z$ to zero simply corresponds to considering
arrays of pancake vortices in perfect registry or,
equivalently, straight vortex lines representing
the crystal\cite{dft}.

The Hansen-Verlet criterion for the freezing
of a two-dimensional liquid\cite{hansen,hanmac}
indicates that freezing occurs when the structure
factor $S(q) = 1/(1- \rho C^{eff}(q))$, evaluated at
$q_m$ {\it i.e.} $S(q_m)$, attains a value of about
5, a value roughly independent of the nature of the
interaction potential. A full density functional calculation yields
the same quasi-universality, a direct consequence of the
fact that correlations, not microscopic potentials,
are the principal determinants of freezing\cite{ry}. Since
C$^{(2)}(q_m) \geq 0$, and C$^{(1)}(q_m)$ is
always reduced (although weakly) in the presence of
disorder, the equilibrium melting line is always {\em
suppressed} by quenched disorder.

We use the following ideas:  (i) The {\em
diagonal} direct correlation function $C^{(1)}(q)$
is weakly affected by disorder and can thus be
approximated by its value in the absence of disorder,
(ii) The off-diagonal direct correlation function
$C^{(2)}(q)$ varies strongly with disorder and
with $H$ and, (iii) $C^{(2)}(q)$
is well approximated at $q=q_m = 2\pi/a$ by its
value at $q=0$.  Since the Hansen-Verlet condition 
is satisfied along the melting line, the following 
holds:
\begin{equation}
\rho_\ell C^{eff} = \rho_\ell(C^{(1)}(q_m) -  C^{(2)}(q_m))\simeq 0.8.
\end{equation}
The off-diagonal correlation function $C^{(2)}$ decays sharply in
in real space; in Fourier space, therefore, its
value at $q = q_{m}$ is close to its value at $q=0$. The validity of
such an  approximation has been tested  in  Ref.~\cite{ankush}. Thus,
\begin{equation}
C^{(2)}(r) \simeq -\beta V^{(2)}(r).
\end{equation}
The prefactor scales with temperature as $\Gamma^2$, implying that 
$\rho_\ell C^{(2)}(q_m) \sim \frac{B}{T^2}$.
Note that $C^{(2)}(q_m)$ increases as $B$ is increased or as $T$ is 
decreased, as is intuitively reasonable.

We observe that $C^{(1)}(q_m)$ increases with a decrease
in $T$; reducing $T$ increases correlations. The
variation in $C^{(1)}$ is expected to be smooth within
the equilibrium disordered liquid. Thus
\begin{equation}
C^{(1)}(q_m;T - \Delta T,B) =  C^{(1)}(q_m;T,B) + q(B,T)\Delta T,
\end{equation}
where $q(B,T)$~$(q > 0)$ is a smooth function of $B$ and $T$ close to the
melting line.  To first order
in $C^{(2)}$, $(B,T)$ can be replaced by
$(B_m,T_m)$ and $C^{(1)}(q_m;B_m,T_m)$ by
its value at freezing for the pure system: $\rho C^{(1)}(q_m;B_m,T_m) 
\simeq 0.8$. We may also neglect the
$B$ and $T$ dependence of $q$ {\it i.e.} $q(B_m,T_m)\simeq q$,
with $q$ a constant; at melting this dependence 
should be small provided $a \ll \lambda$.

An approximate expression for the suppression of the melting line
from its value for the pure case $(B_m,T_m) = (B_m(T),T)$
now follows:  
$\Delta T_m  \sim \frac{B_m(T)}{T^2}$ = $m \frac{B_m(T)}{T^2}$, with
$m$ a constant, related to a temperature derivative of the direct correlation
function. Here $\Delta T_m$ is
the shift in the melting temperature induced by the
disorder.  This relation predicts a larger suppression of the melting
line at higher fields and lower temperatures.

The above result parametrizes the {\em suppression} of the
melting line by quenched disorder. This result can be
combined with results from a calculation of the 
melting line in the pure system to obtain a simple analytic formula 
for the MG-DL phase boundary\cite{mypaper2}. At low fields, a simple Lindemann 
parameter-based calculation of this phase boundary appears to be reasonably
accurate and yields 
$B_m(T) = C(T-T_c)^\alpha$,
where T$_c$ is the critical temperature and $C$ is a constant
appropriate to the pure system\cite{review1}. The exponent $\alpha$ is characteristic of
the fluctuation regime; $\alpha = 2$ describes
to the mean-field case, whereas $\alpha= 1.33$ is
appropriate to the fluctuation dominated $X-Y$ regime.
This then yields the central prediction for the phase boundary $B_m^{dis}(T)$ separating
disordered liquid from multi-domain glass, \cite{mypaper2}
\begin{equation}
B^{dis}_m(T) = C(T -  \frac{m(T-T_c)^\alpha}{T^2} - T_c)^\alpha
\label{eqvgdl}
\end{equation}

In general, we could use any formula here which
best models the melting line in the pure system, as inferred,
for example, from fits close to T$_c$.
Note that the suppression of the pure melting line
by disorder is very weak if $m$ is small. This
suppression becomes progressively large as $m$ is
increased, or alternatively, at a lower temperature
(larger $H$) for given $m$.
It might appear that the final formula (Eq.~\ref{eqvgdl}), with $\alpha,
C$ and $m$ as potential unknowns, contains a fairly
large number of fitting parameters. However, note
that as T $\rightarrow$\, T$_c$, the suppression term
becomes irrelevant. Thus, two of the three parameters
are fixed {\em vis a vis} the pure system, or equivalently,
by an independent fit close to T$_c$; only a single
free parameter is required to fit the {\em suppression} of
the melting line by disorder, provided an independent fit
to the pure system melting line close to H$_{c2}$ is
available.

We now describe the details of our experiments.
The peak effect in pure $2H-NbSe_2$ single crystals
($T_c \simeq 7.1 K$) is tracked by ac susceptibility
(a 10 Hz ac field of amplitude 1 Oe) and dc
magnetization measurements using a Quantum Design SQUID
magnetometer. Susceptibility  measurements are carried out
by varying temperature at different magnetic fields
ranging from 50 Oe - 10000 Oe. Critical currents
are inferred from field dependent magnetization
hysteresis measurements at different temperatures
above 2.5 K. The peak effect manifests itself in a susceptibility 
measurement through a dip in the real part of the ac susceptibility $\chi_1$;
the minimum value of this quantity corresponds to the maximum
value of the critical current. Typical plots of $\chi_1$ vs $T$ 
at applied fields of 6000, 2000 and 200 Oe is displayed in Fig.~\ref{raviacchi}.
Fig.~\ref{criticalcurrent} shows critical currents as inferred from
magnetization measurements as the field is varied. The peak
effect in magnetization measurements is obtained through the
broadening of the magnetization
hysteresis loops. The locus of the peak field
vs temperature H(T) is independent of the technique
used. In both cases a sharp peak is observed.

Fig.~\ref{newdata} contains  the main result of this paper,
exhibiting points for the locus of peak positions
($H_p$,T$_p)$ in $H-T$ space, as obtained from
susceptibility measurements. 
A fit to the freezing line for the multi-domain glass state, Eqn.~\ref{eqvgdl}, is shown in the same plot, with the
fitting parameters $C = 10^4$, $m = 10.0$, $T_c = 7.09$ and
$\alpha =1.33$ (assuming XY exponents). The dashed line is a fit to the case where
$m=0$, as is conventionally assumed in Lindemann-parameter approaches
to the melting transition. Observe the
almost perfect agreement, over 3 decades or more in
magnetic field\cite{ghosh}.
As we argue above, the suppression of the pure melting line can be parametrized
by a single number, related to the suppression of the direct
correlation function in the vicinity of the ordering wave-vector. Thus, our
approach to parameterizing this phase boundary, apart from the agreement with 
experiment we demonstrate, is also economical in terms of being a 
single-parameter fit.

In conclusion,  we stress the main features
of our analysis  and our central result. We compute
the {\em suppression} of the melting phase boundary in the
pure system by quenched disorder.  In contrast to 
approaches based on Lindemann
parameter based measures of the instability of the
Bragg glass phase, we  determine the instability of
the liquid to a static density wave. This instability, within the
density functional formalism,  arises as a
consequence of correlations which build up in the liquid phase;
the method applied here parametrizes the effects of disorder
in {\em reducing} such correlations. In general, it is to be expected that 
theoretical methods
which study the instability towards {\em freezing}
of a fluid phase should better equipped to capture the
physics of the MG-DL line than methods which study
the instability of a solid phase using a Lindemann
criterion or its variants.  

Our central result, an
expression for the disorder-suppressed phase boundary
in the $H-T$ plane across which the disordered
liquid becomes unstable to a state which at least locally
resembles a crystal, provides a remarkably accurate
fit to the experimental data on the classic low-T$_c$ superconductor
2H-NbSe$_2$. The physical picture  we outline is consistent with a general link between the loci of
peak effect phenomena, reflecting the onset of complex dynamics of a driven
vortex system,  and underlying equilibrium order-disorder transitions as mirrored in  the
static phase diagram of Fig.~\ref{phasedia}\cite{mypaper1,mypaper2,mypaper3}. 
More work to test the validity of the theoretical prediction of $B_m^{dis}(T)$ against
data from a variety of superconducting materials as well as to investigate the proposed universality of
the phase diagram of Fig.~\ref{phasedia} would be very useful.

Support from DST (India) is gratefully acknowledged. We thank NEC Research Institute, Princeton, where these
crystals were grown, for support.

\end{document}